\documentclass[5p,sort&compress,lefttitle]{elsarticle} % opcion authoryear para natbib, pero hay que corregir la biblio
\usepackage{graphicx}
\usepackage[left]{lineno}
\usepackage{amssymb,amsmath}
\usepackage{tabularx}
\usepackage[table,xcdraw]{xcolor}

\usepackage[colorlinks=true,linkcolor=blue,urlcolor=blue]{hyperref}

\journal{Ecological modelling}

\begin{document}

\begin{frontmatter} 
\title{A spatially extended model to assess the role of landscape structure on the pollination service of \textit{Apis mellifera}}

\author[ens,cab]{Julien Joseph\corref{cor1}}
\ead{julien.joseph@ens-lyon.fr}
\cortext[cor1]{Corresponding author}

\author[unrn,irna]{Fernanda Santib\'{a}\~{n}ez}
\ead{fsantibanez@unrn.edu.ar}

\author[cab]{Mar\'{\i}a Fabiana Laguna}
\ead{lagunaf@cab.cnea.gov.ar}

\author[cab,ib]{Guillermo Abramson}
\ead{abramson@cab.cnea.gov.ar}

\author[cab,ib]{Marcelo N. Kuperman}
\ead{kuperman@cab.cnea.gov.ar}

\author[unrn,irna]{Lucas A. Garibaldi}
\ead{lgaribaldi@unrn.edu.ar}

\address[ens]{\'Ecole Normale Sup\'erieure de Lyon, Universit\'e Claude Bernard Lyon I, Universit\'e de Lyon, 69342 Lyon Cedex 07, France.}
\address[unrn]{Universidad Nacional de R\'{\i}o Negro. Instituto de Investigaciones en Recursos Naturales, Agroecolog\'{\i}a y Desarrollo Rural. San Carlos de Bariloche, R\'{\i}o Negro, Argentina.}
\address[irna]{Consejo Nacional de Investigaciones Cient\'{\i}ficas y T\'{e}cnicas. Instituto de Investigaciones en Recursos Naturales, Agroecolog\'{\i}a y Desarrollo Rural. San Carlos de Bariloche, R\'{\i}o Negro, Argentina.}
\address[cab]{Centro At\'omico Bariloche (CNEA) and CONICET, R8402AGP Bariloche, Argentina.}
\address[ib]{Instituto Balseiro, Universidad Nacional de Cuyo, R8402AGP Bariloche, Argentina.}

\begin{abstract}

\textit{Apis mellifera} plays a crucial role as pollinator of the majority of crops linked to food production and thus its presence is currently fundamental to our health and survival. The composition and configuration of the landscape in which \textit{Apis mellifera} lives will likely determine the well-being of the hives and the pollination service that their members can provide to the crops. Here we present a spatially explicit model that predicts the spatial distribution of visits by \textit{Apis mellifera} to crops, by simulating daily trips of honey bees, the demographical dynamic of each hive and their honey production. This model goes beyond existing approaches by including 1) a flower resource affected by the feedback interaction between nectar extraction, pollination, blossoming and repeated visits, 2) a pollinators dynamic that allows competition through short term resource depletion, 3) a probabilistic approach of the foraging behavior, modeling the fact that the pollinators have only partial knowledge of the resource on their surroundings, and 4) the specific and systematic foraging behavior and strategies of \textit{Apis mellifera} at the moment of choosing foraging sites, as opposed to those adopted by solitary and wild pollinators. With a balance between simplicity and realism we show the importance of keeping a minimal fraction of natural habitat in an agricultural landscape. We also evaluate the effects of the landscape's structure on  pollination, and demonstrate that there exists an optimal size of natural habitat patches that maximizes the pollination service for a fixed fraction of natural habitat.
\end{abstract}

%%Research highlights (requiere elsart 3.2+)
%\begin{highlights}
%\item Research highlight 1
%\item Research highlight 2
%\end{highlights}


\date{\today}

\end{frontmatter}

%*******************************************************************
\section{Introduction}

The intensive agricultural production systems developed during the twentieth century are facing multiple crises, mostly due to the degradation of the natural ecosystems they  promote~\cite{Newbold,Wilting,Grab}. One of these crises is the fast decline of the populations of pollinators, and the consequent decrease in the quality of the pollination service~\cite{Potts 2010,Vanbergen}. The reduction of floral diversity requested by intensive monoculture particularly endangers wild pollinators, but also weakens managed  species~\cite{Potts 2010,Vanbergen,Potts 2016,Klein}. Pollinators are not only essential for the functioning of  terrestrial ecosystems, but they are also fundamental to our survival~\cite{Potts 2016,Garibaldi 2019}.  Even if recent studies demonstrate the importance of wild bees, the managed bee \textit{Apis mellifera} holds a particularly important place in the pollination service~\cite{Garibaldi 2013,Hung}. 

In this context, countermeasures are constantly being developed to reach a better agriculture: productive, sustainable and healthy for the consumer~\cite{Garibaldi 2019,UNEP,Foley}. These include the structuring of landscapes to improve the ecological services provided by the biological  actors of the crops'  growth and reproduction, such as pollinators~\cite{Garibaldi 2019}. The experimental approach to study the effect of landscape structure on pollination quickly reaches limits in terms of the quantity of available experimental farms and of reproducibility of experiments. An approach in terms of mathematical modeling is particularly fit in such cases, since it allows the fast exploration of a variety of scenarios.

Landscape configurations found in agricultural exploitations around the world have arisen mostly from the experience of the growers, and are usually designed to increase the size of the cultivated area~\cite{MacDonald}. The treatments and management tested on the field or \textit{in silico} to improve the pollination service usually consist in the addition of semi-natural habitat---such as flower strips or grassy bands---mostly on the margins of the crops~\cite{Haussler,Scheper,Jonsson}. But little is known about the global effects of the topology of the spatial arrangements of habitat in agricultural systems~\cite{Martin}.

\begin{table*}[t]
\caption{A summary of the main characteristics of several pollination-foraging models, as discussed in the Introduction.}
\medskip
\newcolumntype{Y}{>{\footnotesize\raggedright\arraybackslash}X} % smally y justificado a izq queda mejor
\begin{tabularx}{\textwidth}{Y|YYYYYY}
\hline
	            & BEEHAVE &	Olsson et al. (2015) & BEESCOUT	& Häussler et al. (2017) & Bumble-BEEHAVE & This work \\ 
\hline
\hline
Species studied	& \textit{A.~mellifera} & Pollinators in general &	\textit{A.~mellifera} & \textit{B. terrestris} & \textit{B. terrestris} & \textit{A.~mellifera} \\
\hline
Level   of organization  & Colony                                                                                   & Community                                                    & Individual                                                                      & Colony                                         & Colony                                                                                                   & Colony                                                                                                      \\
\hline
Spatial   resolution     & $\times$                                                                                 & 25   m / 3 km                                                & Defined by user                                                                 & 25   m / 3 km                                  & 25   m / 5 km                                                                                            & 10   m / 3 km                                                                                               \\
\hline
Pollination   feedback      & $\times$                                                                                 & $\times$                                                     & $\times$                                                                        & $\times$                                       & $\times$                                                                                                 &  \checkmark                                                                                                 \\
\hline
Inter-hive competition & $\times$                                                                                   & $\times$                                                     & $\times$                                                                        & $\times$                                       & \checkmark                                                                                               &  \checkmark                                                                                                 \\
\hline
Site   election modality & based on detection probabilities derived from BEESCOUT and recruitment, which is affected by the energetic efficiency of the trip & Deterministic only based on distance.                        &  Based on simulation of empirical flight patterns & Deterministic, based on resource and distance. & The best patch is chosen out of a probabilistically chosen metapatch, depending on resource and distance. & Each patch has a probability to be chosen based on distance, resource and the resource of neighbor patches. \\
\hline
Goal of the model       & Impact of many stressors on individuals and colony. & Effect of landscape on flower visit rates and bees' fitness. & Understand landscape exploration by bees.                                       & Effect of landscape on flower visit rates.     & Impact of many stressors on individuals, colony, populations and community with mapping.                 & Effect of landscape on flower visit rates.                                                                  \\ 
\hline
\end{tabularx}
\label{tablemodels}
\end{table*}

Many models have been developed in order to understand the needs and implications of sustainable pollination. Some of them consider a homogeneous environment and two interacting populations (plant and pollinator)~\cite{Khoury,Montoya,Cropp}. They generally do not include the spatial dimension, ignoring the effect of plants distribution in space or the distance between plants and pollinators' nests. Nevertheless, such models have provided an understanding of many characteristics of the plant-pollinator interactions in terms of the impact of the presence of semi-natural habitat~\cite{Montoya}, or the coevolutive dynamics between plant and pollinator~\cite{Cropp}. Besides, Croft et al.~(2018)~\cite{croft2018}, motivated by the growing concern on the abrupt and unexplained collapse of honeybee colonies, presented a model to evaluate the impact of pesticides and the regulation of their use on the sustainability of a colony. The authors concluded that the decrease of the colony size is a much better indicator of reduced ecoservice than an increased mortality of brood or foragers.  While the work is focused on a very detailed description of the population dynamics of the individuals within the colony, with special attention on the effect of pesticides in three development stages, it does not incorporate details about the interaction between the bees and the resource. As a general feature, these models are not suitable to study the effect of landscape configurations, as the lack of a spatial dimension conceals the importance of a patterned distribution of resources, where boundaries and heterogeneous distributions can play extremely relevant roles, as will be discussed in the present work.

It is fair to mention, regarding the inclusion of a spatial dimension, several  spatially explicit models that also incorporate seasonal aspects of plants and pollinators dynamics. They are able to predict the probable distribution of the pollinators on a large scale according to the resource distribution~\cite{Lonsdorf,Olsson}. They can also infer the impact of some forms of management or treatments of a particular landscape on pollination~\cite{Haussler}. 
 
In particular, the BEEHAVE family of models has ad\-dressed some questions on pollination using a spatially explicit approach~\cite{Becher 2014,Becher,Becher 2018}. These models addressed the impact of a great number of stressors on pollinators at different scales (individuals, colony, population), while the model we present here focuses on the effect of landscape structure on flower visit rates. Bumble-BEEHAVE is the first model to include competition within and between hives and has shown its capacity to estimate at large scale the flower visit rates~\cite{Becher 2018}.
The study of this last aspect is the nucleus of another very interesting model, which is the one underneath the software tool BEESCOUT~\cite{Becher}. The model implemented in that tool points to examine the exploratory patterns of bees considering the particular structure of the landscape. It includes details about the detection capabilities of bees and propose several search strategies. The aim of this work is to contribute to the simulation of realistic scenarios of colony growth and death in response to environmental and resource changes. In particular, the implementation of the foraging site election in these models leads to a systematic overchoosing of the most resourceful sites inside of a chosen ``metapatch'' of the grid. The sites with few resources are also excluded from the options of the bumblebees. Even if these approximations are useful to save computation time and are reasonable at some level to compare quite different landscapes, suboptimal choices due to partial knowledge of the surroundings and to communication failure between scouts and foragers have been shown to increase the dispersion of the visit and improve the pollination service~\cite{Okada}. Our model uses parallel computation on GPUs (Graphics Processing Units) to avoid these approximations, and propose an adjustable site election modality allowing for mistakes. The scale at which our model operates is also finer (10 m against 25 m for Bumble-BEEHAVE) allowing to explore the question of the size of patches on the pollination efficiency. A comparison of our model to previous foraging models is shown in Table~\ref{tablemodels}.

All the models cited above only consider a resource dynamic that is not affected by pollination~\cite{Lonsdorf, Olsson, Haussler, Becher 2014, Becher, Becher 2018}. This is the second feature present exclusively in our model: a feedback of the resource in response to pollination. Even if, in general, nectar recuperation happens within a day~\cite{Krlevska}, a visited patch is likely to offer less resource on days following a visit because successfully pollinated flowers will stop producing pollen and nectar, and also because some of the flowers can be damaged by over-visitation~\cite{Agustin}. This has the effect of increasing the competition between hives due to a local short-term resource depletion.

Regarding the colony dynamics, our model uses the previously established growth model for \textit{Apis mellifera} of Khoury et al.~(2013)~\cite{Khoury}, combined with a novel foraging model which benefits from recent advances in parallel computation in order to infer the spatial distribution of visits to flowers from a fine resolution map of land cover compatible with a Geographic Information System (GIS). We consider only managed hives of \textit{A.~mellifera} whose well-being only depends on the available resource of their surroundings. We chose to model only managed hives because bee-keeping is a worldwide spread practice and the results we can produce will be applicable to more landscapes. In order to isolate the effects of the landscape structure we also compare the results obtained with the spatially 
explicit model to a similar mean-field model that assumes a uniform distribution of the different types of land covers.

To summarize, in this work we propose a step towards a theoretical foundation for agricultural landscape design, studying the effects of general compositional and structural features, such as the fraction of natural habitat, size of patches, and meshing of the patches of different crops on artificial maps, in order to be able to design new fields that optimize the pollination by \textit{A.~mellifera}. We intend to assess a number of relevant questions, such as: What is the influence of different configurations of natural habitats and crops (considering contrasting edge densities) on the dynamics of the honey-bee population, honey production, and crop flower visitation? How does the number of hives (as a management tool) interact with the effects of the configuration of natural habitats and crops through local competition? And also, what is the influence of different crop blooming periods? In the next section we describe the conceptual and mathematical details of our model, followed by results, a specific mean-field model to contrast them, and a final discussion.

%*******************************************************************
\section{Spatially explicit model}
%\label{model}

In this section we present the spatially explicit model. It is followed in the next section by  a description of the mean field model, to be able to discuss  the  agreements and differences  between them later on.

\subsection{Landscape maps and resource dynamics}

The maps used by the model are grids readable by GIS software, composed of a discrete number of habitat categories. Each habitat has a resource carrying capacity, a daily resource  renewal rate and a blooming period. In this study we will consider that the carrying capacity is constant during the blooming period, and equal to zero the rest of the year, but a temporal distribution of the value of the carrying capacity can also be provided. The resource consists of both nectar and pollen considered together in the present model. Time evolves in discrete steps, with a time unit of one day. 

When the resource is depleted a fraction of it disappears, proportionally to the number of visits, and it is renewed during the following days. Let $K$ be the carrying capacity and $R_t$ the value of the resource before a number $V_t$ of visits on day $t$. Then, the resource of a patch evolves as:
\begin{equation}
\label{eq:resource}
R_{t+1}=R_t-pV_t+r(K-(R_t-pV_t)),
\end{equation}
where $p$ is the fraction of flowers which will cease to produce nectar or pollen after the visits of bees (i.e. the success of pollination plus the damaged flowers), and $r$ is a constant renewal rate. This equation reflects the decrease of the resource due to the daily visits and the recovery dynamics limited by saturation.

For the present study we created random maps of 3 km by 3 km, with each cell measuring 10 m by 10 m. This scale is fine enough to draw realistic maps relevant to the size of \emph{A.~mellifera} foraging sites  (around 30 m by 30 m, i.e. 9 cells), and it is large enough to be interesting regarding their flight range, such that not all hives can forage on the entire map. These maps are composed of two habitats: a natural one whose blooming period is all spring and summer due to the great diversity of flowers it contains, and a monoculture with a 35 days blooming period. We used maps with different  fractions of natural habitat and of edge density (number of cells of a given habitat which are at the border, divided by the total number of cells of the habitat). To build them, we used an algorithm of nucleation in 2D~\cite{Nucle}, varying the number of patches of natural habitat and the number of nuclei from which they grow. More nuclei mean smaller and more dispersed patches, and thus a higher edge density. The value of the resource is the same in the natural habitat and in the cultivated crop, so that the bees do not have any preference when the flowering periods overlap. This is a particular choice, used as a first step in this analysis, but can be easily relaxed when studying specific examples of crops and natural habitats.

\subsection{The hives}

Each hive is characterized by its position on the map, its population of bees and its quantity of honey. When the population of a hive falls below a threshold (one thousand bees),  the hive is considered dead and removed from the list of hives. 

Given that our main interest is the estimation of the number of visits to the crops, the only seasons considered in this study are spring and summer, when the crops bloom. Managed hives are usually well cared for during autumn and winter~\cite{Furgala} and we did not model their dynamic during this season. We only considered that at the beginning of a new foraging season, approximately one third of the bees of each hive has survived.

The dynamic of the hive is based on the one described by Khoury et al.~(2013)~\cite{Khoury}. The only difference is that, while in that model the food increases proportionally to the number of foragers, in our model it increases with the daily gain of the foragers, $G_t$. The new equation for the food dynamic is the following:
\begin{equation}
\label{eq:food}
F_{t+1}=F_t-\gamma_A(n_f+n_w)-\gamma_B n_b+G_t,
\end{equation}
with $F_t$ the quantity of food on day $t$, $\gamma_A$ the consumption rate of workers and foragers, $\gamma_B$ the consumption rate of broods (as in~\cite{Khoury}), and $n_{f,w,b}$ are the number of foragers, workers and broods, respectively.

\textit{Apis mellifera} is faithful to their harvesting sites: when they choose a site, they forage on it all day sending a great number of bees, sometimes until the resource of  nectar is completely drained~\cite{Rollin}. In the model, each day a hive chooses a number of foraging sites with a size of 30~m by 30~m (9 cells), where it will send a fraction of foragers to harvest. The daily gain of the hive is the sum of the gains that each ``squad'' of foragers could obtain from the foraged site:
\begin{equation}
\label{eq:ganancia}
G_t=\sum_{i=1}^{M(B_t)}g_{i,t}(x,y),
\end{equation}
where $M$ is the number of squads which can forage from a hive of size $B_t$ (i.e.~an integral fraction of the number of bees $B_t$), and $g_{i,t}(x,y)$ is the local gain of squad $i$ at  the foraging site centered at $(x,y)$. 

\emph{Apis mellifera} choose their harvesting sites at the beginning of each day relying on relevant information provided by scouts~\cite{Frisch}: the resource at the sites and their  distance from the nest. In the model, we implement this by choosing harvesting sites at random, at the beginning of each day, according to a specific distribution probability as follows. The probability to choose site $(x,y)$ is defined as depending on $R_t(x,y)$, which is the total resource in the 9-cells Moore neighborhood centered at $(x,y)$, and the flight cost to reach them, $f(d)$: 
\begin{equation}
\label{eq:preference function}
P(x,y)= 
\begin{cases}
    \dfrac{(R_t(x,y)f(d))^{\gamma}}{\sum\limits_{(u,v)\in r_f}(R_t(u,v)f(d))^{\gamma}},& \text{if } d<r_f,\\
    0,              & \text{otherwise},
\end{cases}
\
\end{equation}
where $d$ is the Euclidean distance from the hive to $(x,y)$ and $r_f$ is the range of flight of the bees. The cost $f(d)$, normalized to the interval 
$[0,1]$, is an affine decreasing  function of $d$, with $f(0)=1$ and $f(r_f)=0$.
The exponent $\gamma$ represents the knowledge that the hive has of the resource on its surroundings; if $\gamma=0$, the hive chooses its foraging 
sites uniformly at random, and when  $\gamma\to\infty$, the hive systematically chooses the site with the highest harvestable resource. The precise value of this parameter is less important than the general shape of the function, so there is some freedom to choose it within the boundaries of biological significance. In our studies we used $\gamma=3$, which gives a realistic behavior for the scouts, based on the shape of the preference function.

Once the harvesting sites are chosen, the hive sends a fraction $\nu_i$ of foragers to each site $i$, also depending on resource and distance. This 
parameter $\nu_i$ is a weight  factor so that the more harvestable resource there is on a site, more bees are sent to it:
\begin{equation}
\label{eq:ponderation}
\nu_i= \frac{R_t(x,y)f(d)}{\sum\limits_{\text{chosen }(x,y)}R_t(x,y)f(d)}.
\end{equation}

As a result of this election process, several hives can choose the same sites and thus compete for the resource. Note that the probability for hives to choose the same site  is higher  when the resource is globally low on their surroundings, and thus the competition is stronger.

We implement this possibility of hive competition by sorting each day a random harvesting order $\tau$ for each hive. Sorting this order every time step averages the effect  of the chosen order. Let $R_t^\tau(x,y)$ correspond to the value of the resource after the turn of the hive which feeds at round $\tau$. The actual harvestable resource for  the next hive at the site is the product of $R_t^\tau(x,y)$ and $f(d)$. Let $c$ correspond to the maximum that a forager can carry back if the distance to the nest is null. If the resource is in excess, the bees sent to the site harvest the maximum value they can carry during the day, $cf(d) \,\nu_k \,n_f$, but if the resource is lower than this value, they take it all. The daily local gain of hive $i$ at site $(x,y)$ can then be described as follows:
\begin{equation}
\label{eq:ganancia local}
g_{i,t}(x,y)= \max\left(c f(d) \,\nu_i \,n_f , R_t^\tau(x,y) \right).
\end{equation}

In the results we will also discuss the total number of visits, which is a proxy for the pollination efficiency. The number of visits to a site on one day is the sum of the visits  made by all the visiting hives. In turn, the visits made by each of these hives is the product of the number of bees that flew to the site ($N_{trips}$), the average number of flowers visited in one trip ($N_{flower,trip}$) and the average number of trips made by one bee: 
\begin{equation}
\label{eq:visits}
V_t=\sum_{\substack{chosen\\ sites }} N_{trips} N_{flower,trip}\,\nu_i n_f.
\end{equation}

Appropriate values are used for these parameters to compute the pollination service, as will be discussed below.

\subsection{Setting the parameters}

Since our main interest is the impact of landscape structure on pollination, all the parameters which concern the hives dynamic have been kept constant in the simulations (both in the spatially  explicit model and in the mean field presented below). All the parameters used for the colony model have been kept from the study by Khoury et al.~(2013)~\cite{Khoury}.

The parameters of the foraging model have been estimated from real data, such as the number of chosen sites and the number of trips per bee per day, and have been set within the range of observed values. A summary of the parameters and their values is provided in Table~\ref{tabla}.

\begin{figure*}[t]
\includegraphics[width=\textwidth]{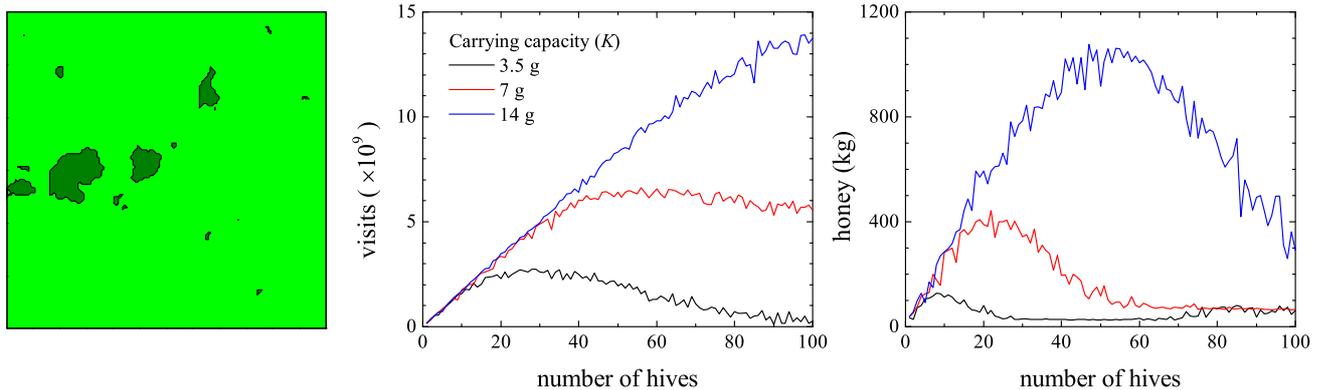}
\caption{The effect of local competition. Left: A synthetic map, with 4\% natural habitat (dark green) and edge density $0.19$. Center: Visits to the crop. Right: Honey production. Each color corresponds to a value of the carrying capacity per cell, as shown. Resource renewal rate, $r$, is 100\% per day. The crop blooming period is from day 20 to day 55, with day 0 the first day of spring. The model ran 2 years, and we took the values for the second year. The corresponding plot for number of bees can be found in the Supplementary Material.}
\label{fig:map90}
\end{figure*}

The average number of trips per day and the number of flowers visited have been chosen as the overall daily number of trips of a hive without distinguishing pollen trip from nectar trip.

The maximum distance of foraging has been set to 3 km so that bees can forage on almost all the map. It is known that bees can make trips farther than 3 km ~\cite{Abou-Shaara} but these trips cannot be seen on the map as currently implemented. The average foraging distance of the bees is not a parameter of the model, but a result of the foraging process, and varies with the exponent $\gamma$ of the preference function and with the distribution of the resource.

The number of visits necessary to pollinate a flower is a parameter difficult to come by as it is highly dependent on the type of flower, and also depends on the scale considered in a nontrivial way. It is not available in the literature for the scale considered here, so we had to adjust it to observe a reasonable behavior of the resource depletion.

A second parameter which is not available in the literature is the size of a foraging squad. Nevertheless, as shown in the sensitivity analysis in the Supplementary Material, its value does not influence much the outputs of the model.

All the codes are written in Python using the CUDA module for GPU computation.

%*******************************************************************
\section{Mean-field model}
\label{mfm}

Let us now describe a mean-field model of the system, which assumes a uniform distribution of the different types of land covers.

The dynamic of the hives during autumn and winter is the same as in the spatially explicit model: bees stop their foraging activities to enter in dormancy. For spring and summer, the mean-field equations for the number of bees in each hive and for the quantity of honey stored in the hive are also the same as in the spatially explicit model (from~\cite{Khoury}). However, the equation for the daily gain $G_t$ differs from Eq.~(\ref{eq:ganancia}), as now  it is proportional to the number of foragers and to the available resource within the flight range of the hive, but has no dependence on the spatial distribution of the resource. In the following we discuss its dynamic in more detail.

As in the spatially explicit model, we consider that the hives are surrounded by two types of habitats: one is a monoculture relying on pollination, while the other is a natural habitat with greater flower diversity, such as a forest or even a semi-natural habitat. The monoculture has a determined blooming period, whereas the natural habitat is flowering during all spring and summer.  We assume that during the blooming period the resource is in excess, and call $R_{max}$  the maximum that a hive can harvest in a day. Out of the blooming period, $R_{bas}<R_{max}$ represents the basal resource available in the natural habitat during all spring and summer. When the fraction of natural habitat is small, a small increment of its value induces a proportional increase of the harvest. But when this fraction is large, the gain does not increase as much with the addition of natural habitat  because it is limited by the foraging capacity of the bees. Let us say that $x$ represents the fraction of natural habitat within the flight range of the hive and $k_s$ is the saturation constant of the harvest. The gain is: 
\begin{equation}
\label{eq:prefer function}
G_t= 
\begin{cases}
    n_f R_{max},& \text{during the crop blooming period,}\\
    n_f R_{bas},& \text{otherwise,}
\end{cases}
\end{equation}
where
\begin{equation}
\label{eq:Rbas}
R_{bas}=R_{max}\dfrac{x(1+k_s)}{x+k_s}.
\end{equation}

The number of visits from the hive to the crop each day is proportional to the number of foragers and to the fraction of monoculture:
\begin{equation}
\label{eq:visit}
V_t=n_f(1-x),
\end{equation}
where $(1-x)$ represents the fraction of crop within the flight range of the hive.

\begin{figure}[t]
\centering
\includegraphics[width=0.8\columnwidth]{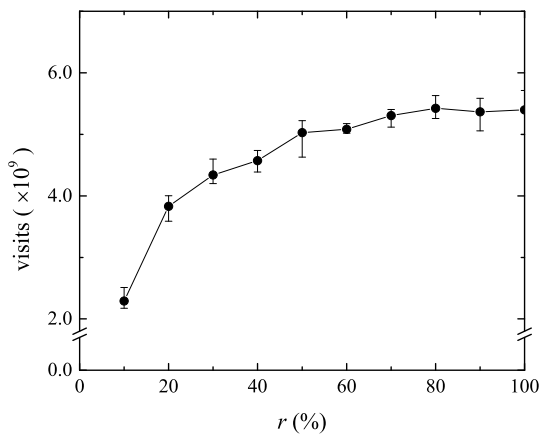}
\caption{Average number of visits as a function of the daily renewal rate of the resource (with minimum and maximum shown as error bars). The carrying capacity of the resource is set as $K=7$ g, and the map is the same as the one used in Fig.~\ref{fig:map90}. The crop blooming period is from day 20 to day 55, with day 0 the first day of spring. The model ran 2 years with 100 hives placed at random on the border of the natural habitat, and we took the values for the second year. The corresponding plots for honey and for number of bees can be found in the Supplementary Material.}
\label{fig:visitsvsr}
\end{figure}

%*******************************************************************
\section{Results}
\label{results}

Let us first explore the effect of competition on the pollination service. In previous models, the absence of competition produces a linear dependence of visits on the number of hives for a given map \citep{Lonsdorf,Olsson,Haussler}. In our model the competition induces a saturation of the number of visits and even a decrease of the visits due to competition-induced hive mortality. We explore how the carrying capacity of the resource and its renewal rate affects this dependence and thus the pollination efficiency regarding the number of hives placed in a crop on two maps (see Fig.~\ref{fig:map90}). We can see that when the resource is in excess and the renewal rate is high, the number of visits is indeed proportional to the number of hives. When the resource is low, we can observe a saturation of the number of visits due to smaller hives, or even because some of the hives did not get through winter. With this analysis we are able to estimate for each map an optimal number of hives for a given carrying capacity and renewal rate of the resources. We can see that the optimal number of hives increases with the amount of natural habitat, the carrying capacity and the renewal rate of the resource, but these effects are nonlinear and thus harder to predict.

In Fig.~\ref{fig:visitsvsr} we can see that the renewal rate of the resource does affect the number of visits only if it is really low, which can happen at the end of the blooming period for instance. We can conclude that most of the competition occurs within a day. The corresponding plots for honey production and number of bees can be found in the Supplementary Material, and they show a similar response.

\begin{figure}[t]
\centering
\includegraphics[width=0.8\columnwidth]{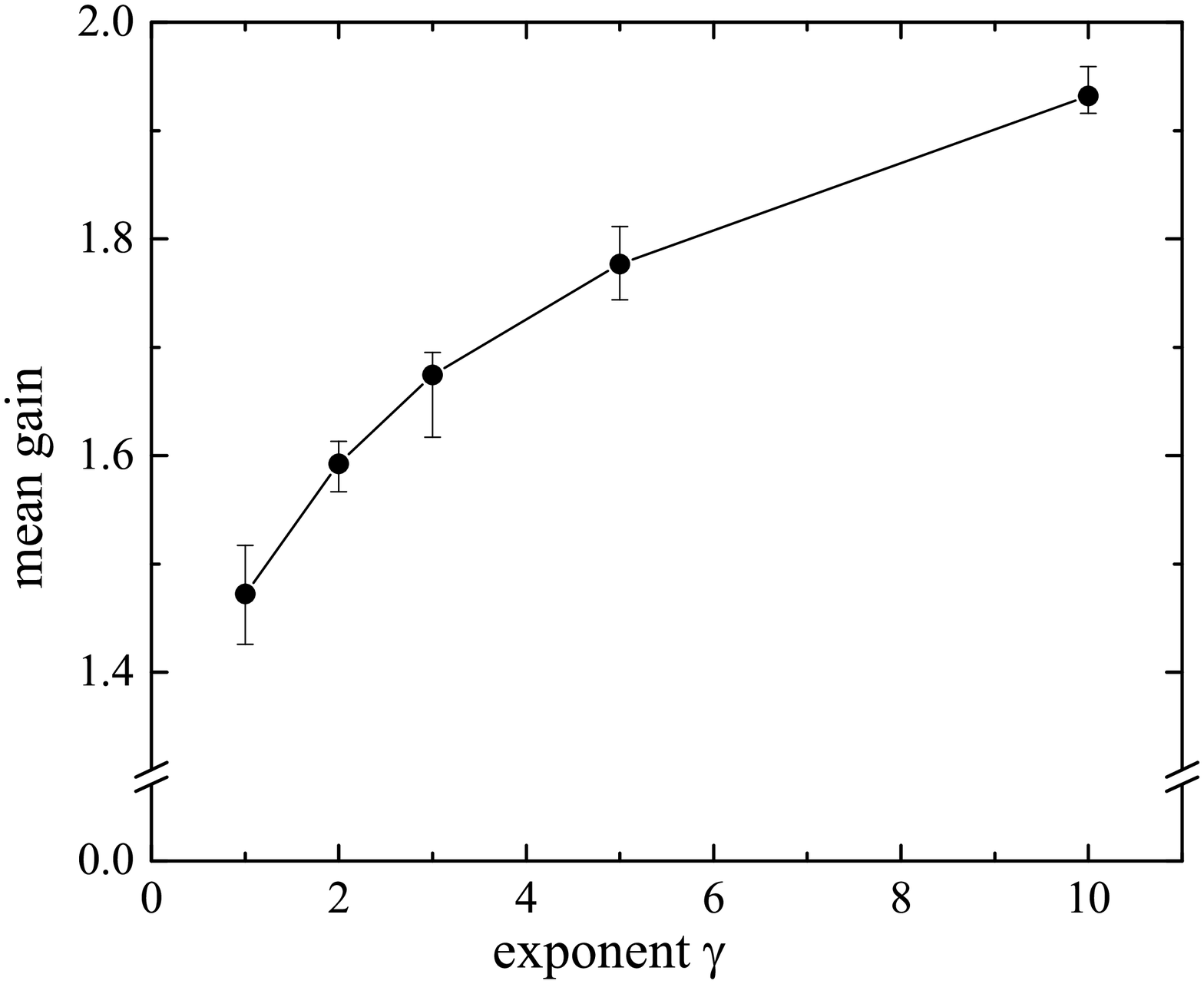}
\includegraphics[width=0.8\columnwidth]{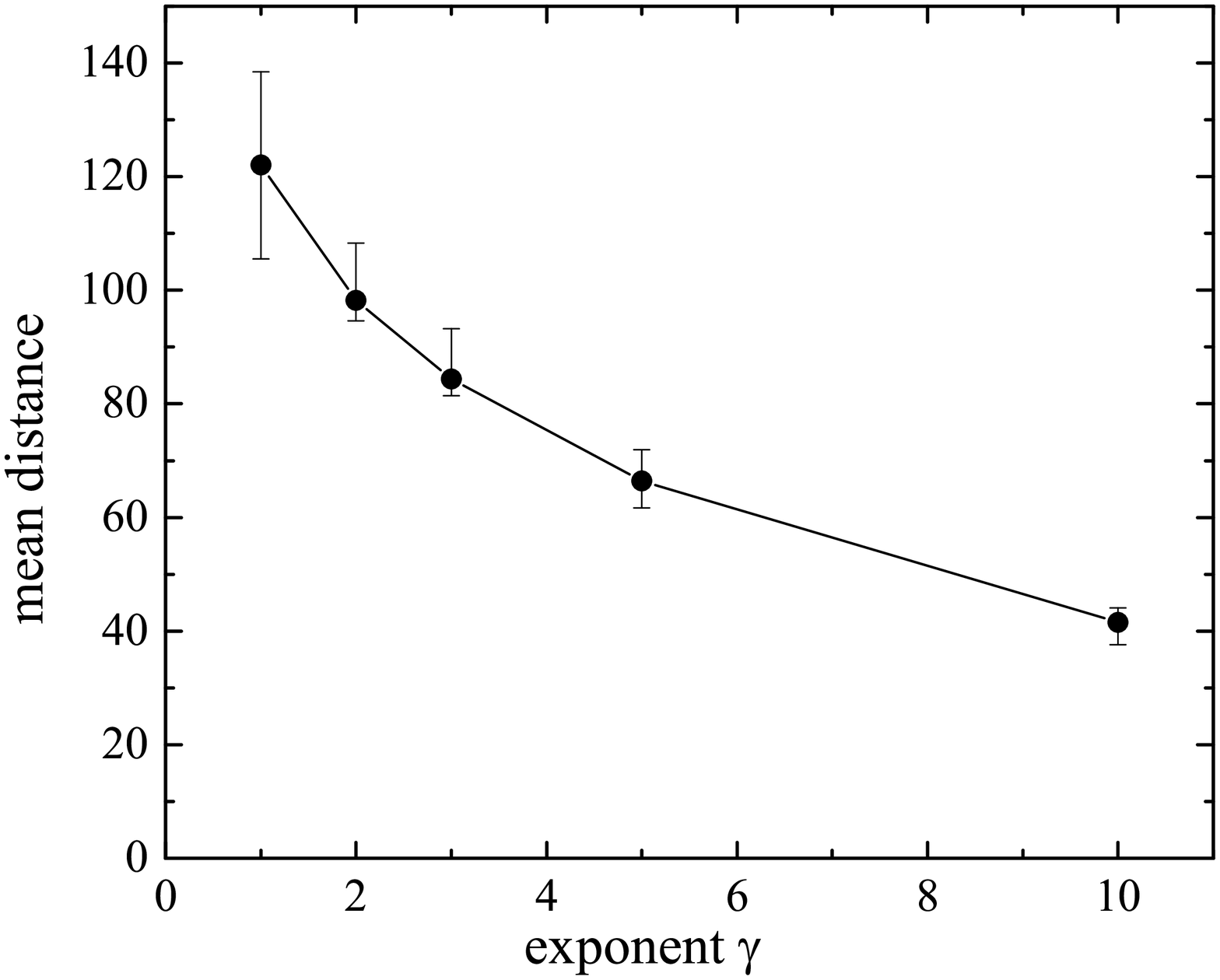}
\caption{Effect of the exponent $\gamma$ of the site election function on
foraging. The mean gain and mean distance are averaged over 10 realizations with a single hive placed at random, and
the minimum and maximum values obtained are displayed as error
bars around the mean. These simulations correspond to a map with
only natural habitat, with carrying capacity of 7 g of resource and 100\% resource
renewal rate.}
\label{fig:gamma}
\end{figure}

We also explored the effect of the exponent $\gamma$ of the site election function (Eq. (\ref{eq:preference function})) on the mean foraging distance of bees and the mean gain of nectar and pollen per site. The results are shown in Fig.~\ref{fig:gamma}. We can see that the exponent performs as expected, increasing the foraging efficiency of the hives and reducing the foraging distance.

\begin{figure*}[t]  
\centering  
 \includegraphics[width=\textwidth]{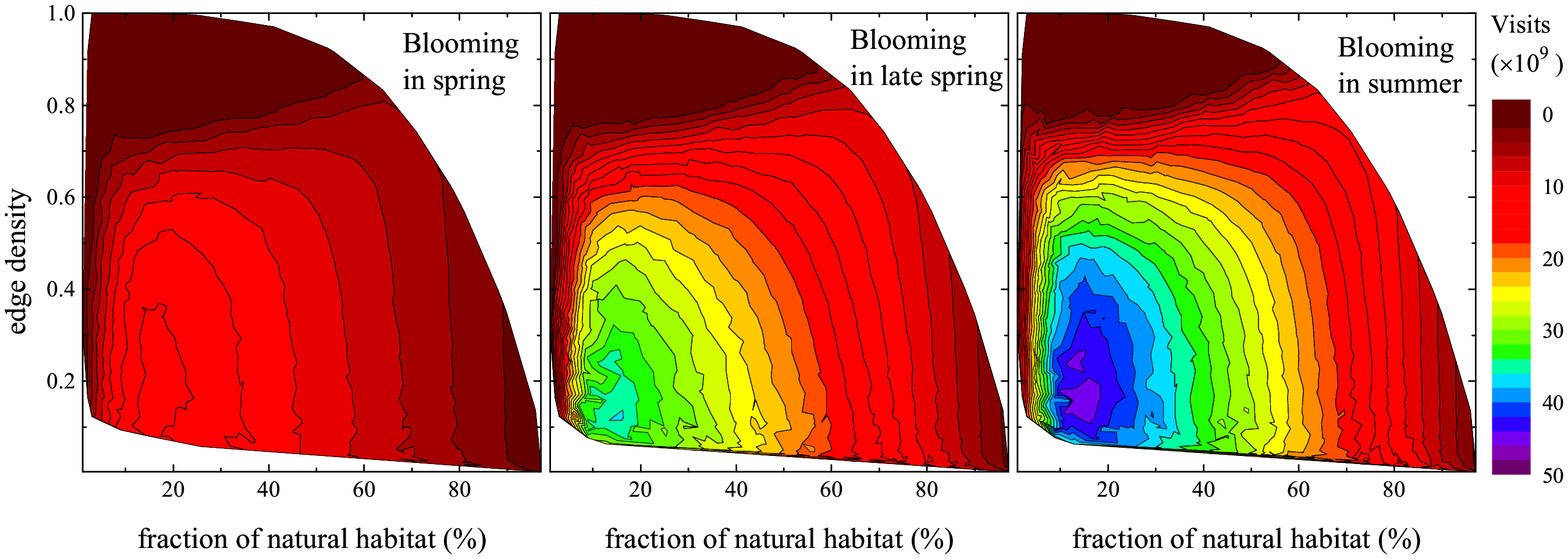}
\caption{Mean annual number of visits to the crop as a function of the fraction of natural habitat and the edge density. The color scale shows the mean annual  total number of visits. Left: crop blooming period from day 22 to day 55. Middle: crop blooming period from day 80 to day 115. Right: crop blooming period from day 130 to day 165. Day 0 is the first day of spring. We used 1000 maps for each of these analysis. Day 0 is the first day of spring. The corresponding plots for honey production and number of bees can be found in the Supplementary Material.}
    \label{fig:visits}
\end{figure*}

To quantify the effect of the amount of natural habitat we performed extensive simulations, monitoring the foraging activity of the bees on one thousand random maps during 4 years. For each map, we placed 100 nests at random on the borders of the natural habitat (which is the usual practice for managed hives). Due to runtime constraints, each map was simulated once. The number of maps was chosen so that the effect of the stochasticity in the heterogeneity of the landscape does not affect significantly the average results. The mean anual number of visits to the crop is shown in Fig.~\ref{fig:visits} for three possible blooming periods during the season. The corresponding contour plots for honey production and number of bees can be found in the Supplementary Material.

The number of visits to the crop reaches a maximum (darkest green in the plots) that remains stable for all the blooming periods (only its value increases when we go later in the season). We can also see that there is an additional dependence on the edge density of the natural habitat with a maximum between 0.2 and 0.4. A low edge density means that the natural habitat and the
crop are well segregated, making some sites at the center
of the crop unreachable for the bees. On the other hand, a
large edge density means that the natural habitat and the
crop are well mixed, diluting the resource of the natural
habitat out of the blooming period. For the number of
visits to be maximal, the patches of natural habitat can
not be too dispersed, but cannot be extremely segregated either.

\subsection*{Results of the mean field model}

\begin{figure}[ht]
 \centering
 \includegraphics[width=0.8\columnwidth]{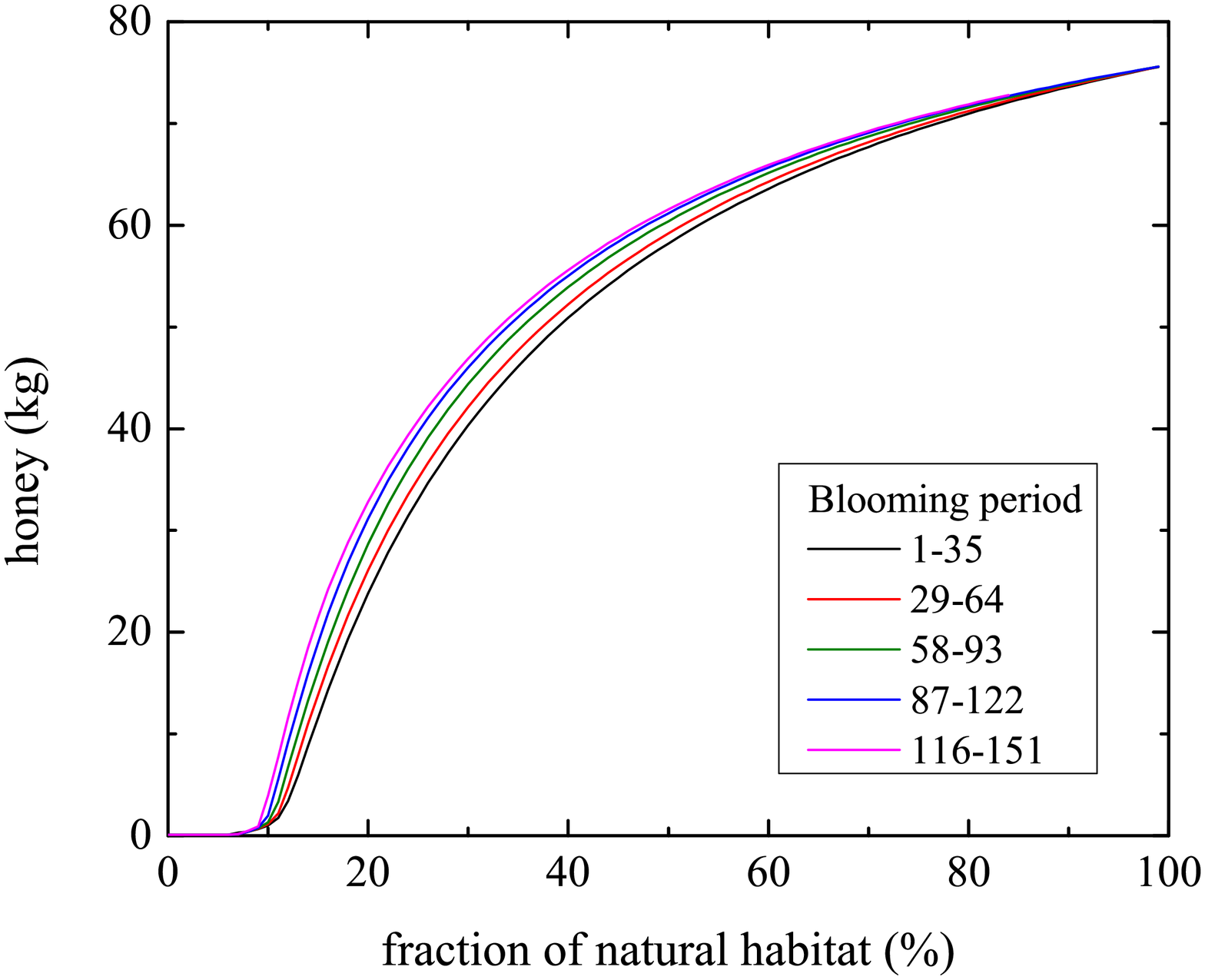}
 \includegraphics[width=0.8\columnwidth]{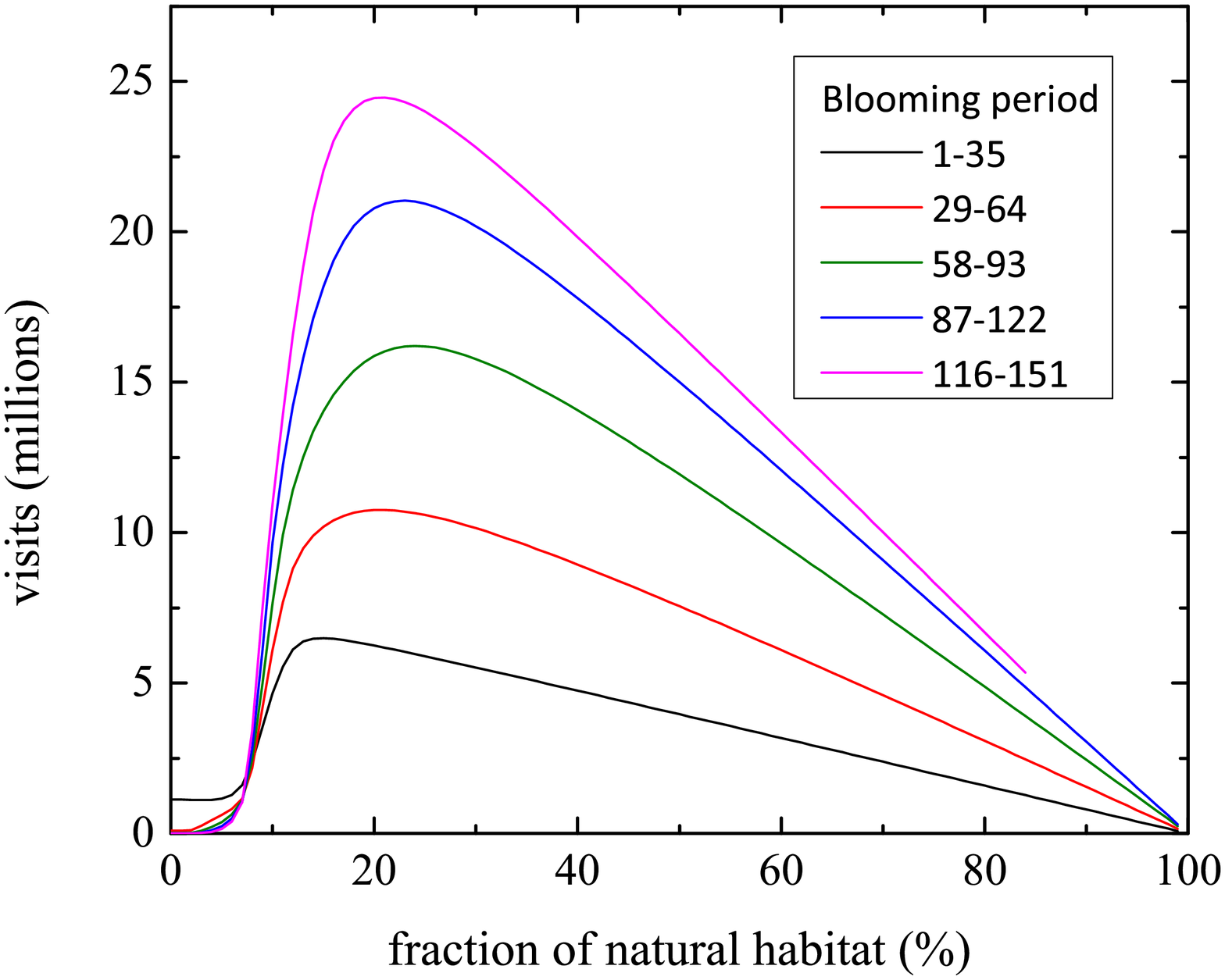}
  \caption{Analysis of the impact of the fraction of natural habitat on the honey production and the annual number of visits with the mean field model.  Top: Mean annual production of honey as a function of the fraction of natural habitat. Bottom: Annual number of visits to the crop as a function of the fraction of natural habitat. The colors correspond to different blooming periods, as shown in the legend in days. Day 0 corresponds to the first day of spring.}
\label{fig:meanfield}
\end{figure}

As mentioned before, a mean field model is suitable to  estimate the influence of the fraction of natural habitat on the pollination service by 
\textit{A.~mellifera} without considering the effect of the distribution and shape adopted by this fraction. Here we present  the corresponding results.  We made simulations for different values of the fraction of natural 
habitat, and measured the number of visits of a hive during a 4 years period. The results are shown 
in Fig.~\ref{fig:meanfield}.

We can see that the production of honey increases with the fraction of natural habitat, saturating for large values of it. We can also see that the  honey production barely depends on the blooming period of the monoculture. The number of visits, instead, strongly depends on the blooming period. It has a maximum for a fraction of natural habitat which ranges from 11.6\% for an early blooming to 28.1\% for a late one. The numbers of visits on late blooming crops are larger because the honey bee population is also larger during this time of the year.

%*******************************************************************
\section{Discussion}

The spatially explicit model presented in this work provides a valuable insight about the impact of the landscape structure on the pollination service.   We have seen that a mean field approach accurately predicts the impact of the fraction of natural habitat on the number of visits. Nevertheless, it fails to explain much of the richness of the possible outcomes since the disposition of the natural habitat patches and the competition between hives showed to have an influence on the pollination service. Even if the existence of an optimal fraction of natural habitat to maximize the number of visits is quite predictable with simpler mean field models or good sense, the value it takes in function of the configuration of the landscape, or the growth of hives remains quite non trivial. It is remarkable that some published experimental results suggest a maximum of pollination around the same values of edge density and fraction of natural habitat as the ones found by means of our model with default parameters (20\% of semi-natural habitat for an edge density between 0.3 and 0.4)~\cite{Martin}. Moreover, this is, as far as we know, the only model which can effectively address the effect of the size of patches on the number of visits considering a cell size smaller than the size of foraging sites, allowing to find the explicit dependence on patch density. We could infer from extensive computations how this optimum varies with forager mortality or flight range of the bees, for instance. The model also showed that the moment of the flowering period does not influence much the optimal composition or configuration of the agricultural landscape for pollination.

Our model does not allow to predict an absolutely trustworthy number of visits. Indeed, several  parameters (such as the number of trips per day and the number of flowers visited per travel), which have been considered constant, are in fact highly variable depending on parameters we chose not to investigate in this first study (weather, subspecies of bees or type of crop, for instance). But taking into account that these parameters vary evenly in every map, we can consider that the model is suitable to compare the number of visits between maps. Specific implementations for real world applications are under way, and will be the subject of further work.

The partial knowledge and the visibility of bees is hard to model in simple mathematical terms. That is why the site election modality of this foraging model, even if it appears less biased than in Bumble-BEEHAVE, remains an approximation. It clearly does not replace approaches based on individual decision making like the BEESCOUT to understand the foraging behavior of bees. But for the moment the computation time required to apply the decision making and field exploration to as much individuals as we have in a 3 km by 3 km map remains too high, and our function remains a good approximation. As it is shown in Fig.~\ref{fig:gamma}, the exponent $\gamma$ can easily be deduced from the mean daily gain of foragers and the mean flight distance of the bees, and allows the user to avoid a quite unrealistic systematic choice of the most resourceful patch, leading to an overestimated competition between foragers for a single patch.

Our results apply to hives which are managed at low cost without intervention during spring and summer. In some places, with a particular lack of natural habitat offering flower diversity, agreements are made between farmers and beekeepers in order to keep alive enough hives for pollination purposes~\cite{Goodwin}. They can be moved across the fields through the seasons, and regularly supplied with syrup if needed. A good beekeeper can effectively erase the effect of landscape structure, but at a great cost in time and money. 

In our analysis we have presented the total number of visits over the map as one of the important outputs, which is different from the economically more relevant fruit production. It is known that generally around ten visits per flower are enough for the fecundation to occur, and that more visits not only do not improve production, but can even damage the pistil and reduce fecundation probability~\cite{Rollin,Agustin}. We have also calculated production from the local number of visits each day, but because hives have identical foraging behavior, fruit production and number of visits are similar. Consequently, we did not find it useful to show these results here. However, we can say that the slight differences observed between visits and production only occur when the competition is strong and the flowers are overvisited. These differences are likely to increase for a larger number of hives. A detailed analysis will also be presented elsewhere.

It is important to rise a word of caution about our results concerning the impact of the fraction of natural habitat on production. What we demonstrated is that there is an optimal fraction  of natural habitat needed in order to maintain a sufficient population of \textit{A.~mellifera} to pollinate the field. But it does not mean that increasing the fraction of natural habitat in an agricultural landscape will necessarily diminish the production. Indeed, a natural habitat such as a forest not only serves as a nectar supply for \textit{A.~mellifera} all over spring and summer, but also provides several other ecological services, e.g.~as a wind breaker, air cooler, furnishing water retention and shelter for wild pollinators~\cite{DeMarco}, etc. The minimal fraction of natural habitat which begins to damage the pollination service by \textit{A.~mellifera} is likely to be lower than the one that will damage the actual yield of crop. 

It is generally acknowledged that, while \textit{A.~mellifera} is the most used pollinator of crops, they are not the most efficient ones~\cite{Garibaldi 2013,Rollin}. A model of pollination by wild bees  using the same resource maps has already been developed by us, and will be used to characterize the effects of the landscape structure on the pollination by native pollinators on artificial and real landscapes, as well as the impact of mixing native and managed bees on pollination.

The goal reached by this work is a characterization of the effect of some structural features of landscapes on the pollination service. The model has been developed to analyze the pollination service in real GIS maps with a larger number of land covers and it is also able to predict the impact of slight changes in real landscape composition and configuration on the pollination service. It will be used in a subsequent study to predict the pollination service on different real agricultural landscapes. We hope that it can become a useful tool for farmers, beekeepers and policy-makers in order to understand the impact of the composition and configuration of the agricultural landscape on the pollination service by the honey bee, help them to design their farms and contribute to a sustainable use of the environment.

%******************************************************************* 
\section{Acknowledgements} We acknowledge financial support several sources:  FONCYT, Plan Argentina Innovadora 2020: PICT 2016-0305, UNRN: PI 40-B-567, UNCuyo: SIIP 06/C546, CONICET: PIP 112-2017-0100008CO.
We also thank the ENS de Lyon  which allowed the visit of J.~Joseph to Bariloche through their internship program. We also thank the reviewers of the manuscript for their valuable comments and suggestions. 

The funding agencies were not involved in the research or the preparation of 
the manuscript.  

\begin{table*}[t]
\caption{Summary of the main parameters used in the model. Other parameters have been set such that the results fit into a biologically relevant range of values for the number of bees and the production of honey: between 0 and 70,000 bees per hive, and between 0 and 50 kg of honey per hive, at the end of summer. Population dynamic parameters differ for each particular climate. We used a set of parameters that corresponds to the region of R\'{\i}o Negro (Argentina). Hive dynamics has been modeled following \cite{Khoury}, with parameters and a sensitivity analysis provided in the Supplementary Material.}
\label{tabla}
\medskip
\begin{tabularx}{\textwidth}{Xcccl}
\hline
Description      & Symbol & Value & Units & Source \\
\hline \hline
Map size  &  & 3  & km  & arbitrary \\
Cell size (harvest site is $3\times 3$ cells)&  & 10 & m   & typical \\
Time step &  & 1  & day & arbitrary \\
Blooming period of the monoculture&  & 35 & days & \cite{Lesser} \\
Flowers per cell	&  &	15,000 &  &	\cite{Lesser} \\
Resource carrying capacity per cell	& $K$	& 3.5, 7, 14 & g	& \cite{Krlevska,Lesser} \\
Daily renewal rate of the resource  & $r$ & 70\%-100\% & & adjusted \\
Number of visits to pollinate a flower	&  &	10	&  & from expertise \\
Resource lost per cell and visit (pollination or flower damage) & $p$ & $4.7\times\!10^{-5}$ & g & derived \\
Knowledge of the environment & $\gamma$ & 3 &  & adjusted \\
Size of a squad	& $k$	& 200	&  & arbitrary \\
Trips per forager per day	& $N_{trips}$ &	19 &  & \cite{Hagler} \\
Flowers visited in one forager's trip &	$N_{flower,trip}$	& 75 & &	\cite{Hagler} \\
Maximum foraging distance	& $r_f$ & 	3 & km	& \cite{Abou-Shaara} \\
Maximum food a forager can bring back on one trip  & $c$ & 0.1 & g & \cite{Khoury} \\
\hline
\end{tabularx}
\end{table*}

%*******************************************************************


\begin{thebibliography}{99}

% agregar [Autor, a\~{n}o] para usar natbib con nombre y a\~{n}o
\bibitem{Newbold} Newbold T et al. (2015). \emph{Global effects of land use on local terrestrial biodiversity}. Nature 520, 45-50.

\bibitem{Wilting} Wilting HC, Schipper AM, Bakkenes M, Meijer JR and Huijbregts MAJ  (2017). \emph{Quantifying biodiversity losses due to human consumption: a global-scale footprint analysis}. Environ. Sci. Technol. 51, 3298-3306.

\bibitem{Grab} Grab H, Bramstetter MG, Amon N et al. (2019). \emph{Agriculturally dominated landscapes reduce bee phylogenetic diversity and pollination services}. Science 363, 282-284.

\bibitem{Potts 2010} Potts SG et al. (2010). \emph{Global pollinator declines: trends, impacts and drivers.} Trends Ecol. Evol. 25, 345-353 .

\bibitem{Vanbergen} Vanbergen AJ and The Insect Pollinators Initiative (2013). \emph{Threats to an ecosystem service: pressures on pollinators}. Front. Ecol. Environ 11, 251-259.

\bibitem{Potts 2016} Potts SG et al. (2016). \emph{Safeguarding pollinators and their values to human well-being}. Nature 540, 220-229.

\bibitem{Klein} Klein AM et al. (2007). \emph{Importance of pollinators in changing landscapes for world crops}. Proc. R. Soc. B 274, 303-313.

\bibitem{Garibaldi 2019} Garibaldi LA, P\'{e}rez-M\'{e}ndez N, Garratt MPD, Gemmill-Herren B, Miguez FE, Dicks LV (2019). \emph{Policies for ecological intensification of crop production}. Trends in Ecology and Evolution 34, 282-286.

\bibitem{Garibaldi 2013} Garibaldi LA, Steffan-Dewenter I, Winfree R, Aizen MA, Bommarco R, Cunningham SA, Kremen C, et al. (2013). \emph{Wild pollinators enhance fruit set of crops regardless of honey-bee abundance}. Science 339, 1608-1611.

\bibitem{Hung} Hung K-LJ, Kingston JM, Albrecht M, Holway DA, Kohn JR. (2018). \emph{The worldwide importance of honey bees as pollinators in natural habitats}. Proc. Biol. Sci. 285(1870), 20172140.

\bibitem{UNEP} Convention on Biological Diversity (CBD) (2000). UNEP Decisions Adopted by the Conference of the Parties to the Convention on Biological Diversity at its Fifth Meeting (UNEP/CBD/COP/5/23/Annex III), Decision V/5 (Nairobi); \url{https://www.cbd.int/doc/decisions/COP-05-dec-en.pdf}.

\bibitem{Foley} Foley, JA et al. (2011). \emph{Solutions for a cultivated planet}. Nature 478, 337-342.

\bibitem{MacDonald} MacDonald JM, Korb P, Hoppe RA (2013). \emph{Farm size and the organization of U.S. crop farming}. Economic Research Report Number 152.

\bibitem{Haussler} Häussler J, Sahlin U, Baey C, Smith HG, Clough Y. (2017). \emph{Pollinator population size and pollination ecosystem service responses to enhancing floral and nesting resources}. Ecol. Evol. 7, 1898-1908.

\bibitem{Scheper} Scheper J, Bommarco R, Holzschuh A, Potts SG, Riedinger V, Roberts SP et al. (2015). \emph{Local and landscape-level floral resources explain effects of wildflower strips on  wild bees across four European countries}. J. Appl. Ecology 52, 1165-1175.

\bibitem{Jonsson}
Jönsson AM, Ekroos J, Dänhardt J, Andersson GK, Olsson O and Smith HG (2015). \emph{Sown flower strips in southern Sweden increase abundances  of wild  bees  and  hoverflies  in  the  wider  landscape}. Biological  Conservation 184, 51-58.

\bibitem{Martin} Martin EA et al. (2019). \emph{The interplay of landscape composition and configuration: new pathways to manage functional biodiversity and agroecosystem services across Europe}. Ecology Letters 22, 1083–1094.

\bibitem{Montoya} Montoya D, Haegeman B, Gaba S, de Mazancourt C, Bretagnolle V, Loreau M. (2019). \emph{Trade-offs in the provisioning and stability of ecosystem services in agroecosystems}. Ecological Applications 29(2), e01853.

\bibitem{Cropp} Cropp R and Norbury J (2018). \emph{Simulating eco-evolutionary processes in an obligate pollination model with a genetic algorithm}. Bull. Math. Biol. 81, 4803-4820.
%\url{https://doi.org/10.1007/s11538-018-0508-1}.

\bibitem{Khoury} Khoury DS, Barron AB, Myerscough MR (2013). \emph{Modelling food and population dynamics in honey bee colonies}. PLoS ONE 8(5): e59084. 
%https://doi.org/10.1371/journal.pone.0059084

\bibitem{croft2018} Croft S, Brown M, Wilkins S, Hart A, Smith GC (2018). \emph{Evaluating European Food Safety Authority protection goals for honeybees (\emph{Apis mellifera}): What do they mean for pollination?} Integrated environmental assessment and management 14, 750-758.
%doi: 10.1002/ieam.4078

\bibitem{Lonsdorf} Lonsdorf E, Kremen C, Ricketts T, Winfree R, Williams N and Greenleaf SS. (2009). \emph{Modelling pollination services across agricultural landscapes}. Annals of Botany 1, 12.

\bibitem{Olsson} Olsson O, Bolin A, Smith H and Lonsdorf E. (2015). \emph{Modeling pollinating bee visitation rates in heterogeneous landscapes from foraging theory}. Ecol. Model. 316, 133-143.

\bibitem{Becher 2014} Becher MA, Grimm V, Thorbek P, Horn J, Kennedy PJ and Osborne, JL (2014). \emph{BEEHAVE: a systems model of honeybee colony dynamics and foraging to explore  multifactorial causes of colony failure}. J. Appl. Ecol. 51, 470-482.

\bibitem{Becher} Becher MA, Grimm V, Knapp J, Horn J, Twiston-Davies G and Osborne JL (2016). \emph{BEESCOUT: A model of bee scouting behaviour and a software tool for characterizing nectar/pollen landscapes for BEEHAVE}. Ecol. Model. 340, 126- 133.

\bibitem{Becher 2018} Becher, M. A., Twiston-Davies, G., Penny, T. D., Goulson, D., Rotheray, E. L., Osborne, J. L. (2018). \emph{Bumble-BEEHAVE: a systems model for exploring multifactorial causes of bumblebee decline at individual, colony, population and community level}. J. Appl. Ecol. 55, 2790-2801.

\bibitem{Okada} Okada, R., Ikeno, H., Kimura, T. et al. (2015). \emph{Error in the honeybee waggle dance improves foraging flexibility}. Sci. Rep. 4, 4175. 
%https://doi.org/10.1038/srep04175

\bibitem{Krlevska} Krlevska H., Kiprijanovski M. and Naumovski M. (1995). \emph{Research on nectar-bearing capacity of apples}. Macedonian Agricult. Rev. 42, 115-118.

\bibitem{Agustin}  S\'aez A, Morales LC, Ramos LY and Aizen AM (2014), \emph{Extremely frequent bee visits increase pollen deposition but reduce drupelet set in raspberry}. J. Appl. Ecol. 51, 1603-1612.

\bibitem{Nucle} Kashchiev D. (2000). \emph{Nucleation: Basic theory with applications} (Butterworth-Heinemann, Oxford).

\bibitem{Furgala} Furgala B. (1975). \emph{Fall management and the wintering of productive colonies}. In: \emph{The hive and the honeybee}, ch. XVI. Dept. of Entomology, University of Minnesota, Dadant.

\bibitem{Rollin} Rollin O and Garibaldi LA (2019). \emph{Impacts of honey bee density on crop yield: A meta-analysis}. J. Appl. Ecol. 56, 1152-1163.

\bibitem{Frisch} von Frisch K (1967). \emph{The dance language and orientation of bees} (Harvard University Press, Cambridge, MA).

\bibitem{Abou-Shaara} Abou-Shaara, H. F (2014). \emph{The foraging behaviour of honey bees, \emph{Apis mellifera}: a review}. Vet. Med. (Praha) 59, 1–10.

\bibitem{Goodwin} Goodwin M (2012). \emph{Pollination of crops in Australia and New Zealand}. Rural Industries Research and Development Corporation Publication No. 12/059.

\bibitem{DeMarco} De Marco P and Coelho FM. (2004). \emph{Services performed by the ecosystem: forest remnants influence agricultural cultures' pollination and production}, Biodiversity and Conservation  13, 1245.

\bibitem{Quebec} Conseil des Productions V\'{e}g\'{e}tales du Qu\'{e}bec (1977). \emph{Apiculture: Emplacement du rucher et d\'{e}rive}. Agdex 616, 4pp.
%https://www.agrireseau.net/apiculture/documents/64092/emplacement-du-rucher-et-derive

\bibitem{Lesser} Lesser Preuss, R (2004). \emph{Manual de apicultura moderna} (Editorial Universitaria, Santiago de Chile). 

\bibitem{Hagler} Hagler JR, Mueller S, Teuber LR, Machtley SA, Van Deynze A (2011). \emph{Foraging range of honey bees, \emph{Apis mellifera}, in alfalfa seed production fields}. J. Insect Sci. 11, 144. 
%doi: 10.1673/031.011.14401. 

%- Gurney, W. S. C., Nisbet, R. M., 1998. Ecological Dynamics. Oxford University Press, New York, Oxford
%- Rosenzweig, M. L., 1971. Paradox of enrichment: destabilization of exploitation ecosystems in ecological time. Science 171, 385-387.
%- Thompson, J., Stewart, H., 2002. Nonlinear Dynamics and Chaos, 2nd Edition. John Wiley and Sons, Chichester, UK.

%\bibitem{Rousseau} Rousseau-Gueutin M, Morice J, Coriton O, Huteau V, Trotoux G, Nègre S, Falentin C et al. (2017). \emph{The impact of open pollination on the structural evolutionary dynamics Allotetraploid Brassica napus L. Genes}.  Genomes, Genet. 7, 705-717.

%\bibitem{Teixeira} Teixeira S, Foerster K, Bernasconi G (2009). \emph{Evidence for inbreeding depression and post-pollination selection against inbreeding in the dioecious plant Silene latifolia}. Heredity (Edinb). 102(2), 101-12.

%\bibitem{Halasz} Hal\'asz  J, Hegedûs A and Pedryc A (2006). \emph{Review of the molecular background of self-incompatibility in rosaceous fruit trees}. International J. Hort. Sci. 12(2), 7-18.

%\bibitem{bolin18} Bolin A, Smith HG, Lonsdorf EV, Olsson O (2018). \emph{Scale-dependent foraging tradeoff allows competitive coexistence.} Oikos 127, 1575-1585.

%\bibitem{Free} Free J (1965). \emph{The behaviour of honeybee foragers when their colonies are fed sugarsyrup}. Journal of Apicultural Research 4, 85-88.

\end{thebibliography}
\end{document}